\documentclass{article}
\usepackage{geometry}
\usepackage[utf8]{inputenc}
\usepackage{cite}
\usepackage{url}
\usepackage{amssymb,amsmath,amsthm}
\usepackage{bm}
\usepackage{graphicx}
\usepackage{enumerate}
\usepackage{microtype}
\usepackage{color}
\usepackage{hyperref}
\usepackage[ruled,linesnumbered,vlined]{algorithm2e}
\usepackage{multirow}

\title{Shape analysis via inconsistent surface registration}
\author{Gary P. T. Choi$^{1}$, Di Qiu$^{2}$, Lok Ming Lui$^{2\ast}$\\
\\
\footnotesize{$^{1}$John A. Paulson School of Engineering and Applied Sciences, Harvard University, Cambridge, MA, USA}\\
\footnotesize{$^{2}$Department of Mathematics, The Chinese University of Hong Kong, Hong Kong}\\
\footnotesize{$^\ast$Corresponding author; E-mail: lmlui@math.cuhk.edu.hk}
}
\date{\today}

\begin{document}

\maketitle

\begin{abstract}
In this work, we develop a framework for shape analysis using inconsistent surface mapping. Traditional landmark-based geometric morphometrics methods suffer from the limited degrees of freedom, while most of the more advanced non-rigid surface mapping methods rely on a strong assumption of the global consistency of two surfaces. From a practical point of view, given two anatomical surfaces with prominent feature landmarks, it is more desirable to have a method that automatically detects the most relevant parts of the two surfaces and finds the optimal landmark-matching alignment between those parts, without assuming any global 1-1 correspondence between the two surfaces. Our method is capable of solving this problem using inconsistent surface registration based on quasi-conformal theory. It further enables us to quantify the dissimilarity of two shapes using quasi-conformal distortion and differences in mean and Gaussian curvatures, thereby providing a natural way for shape classification. Experiments on Platyrrhine molars demonstrate the effectiveness of our method and shed light on the interplay between function and shape in nature. 
\end{abstract}

\section{Introduction}
The use of mathematical mappings for quantifying shape variation began a century ago with the seminal work of D'Arcy Thompson~\cite{Thompson1917on}, in which a theory of transformations was proposed for studying planar projections of biological shapes, such as fish and human skull, with the aid of a deformed underlying grid. Given two sets of corresponding landmarks representing two shapes, traditional geometric morphometrics methods such as the Procrustes superimposition method~\cite{Gower1975generalized} use rigid (translational and rotational), isotropic/anisotropic scaling, and shear transformations to align them for quantifying their shape difference. Because of the limited degrees of freedom of these transformations, more advanced non-rigid mappings such as the thin plate splines method~\cite{Bookstein1989principal} have been considered. In general, any mapping between two surfaces will unavoidably induce distortion in angle or area (or both). There has been a vast number of works aiming to achieve angle-preserving (conformal)~\cite{Haker2000conformal,Levy2002least,Gu2004genus,Crowdy2005the,Crowdy2006conformal,Mullen2008spectral,Lai2014folding,Choi2015fast,Yueh2017an,Choi2018a,Sawhney2018boundary} and area-preserving (authalic)~\cite{Zhu2003area,Zou2011authalic,Zhao2013area,Choi2018density,Yueh2019a,Pumarola20193dpeople,Choi2020area} mappings (see~\cite{Floater2005surface,Hormann2007mesh,Gu2011numerical} for detailed surveys on the subject). These methods, however, do not allow for an exact matching of the corresponding landmarks on two surfaces and hence hinder the accurate quantification of shape variation.

In recent years, quasi-conformal theory has emerged as a useful tool for geometry processing and medical imaging~\cite{Weber2012computing,Lipman2012bounded,Lui2013texture,Choi2016spherical,Choi2016fast,Choi2017conformal,Naitsat2018geometric,Choi2019parallelizable}. In particular, landmark-matching quasi-conformal mapping methods have been developed for image and surface registration~\cite{Lam2014landmark,Yung2018efficient,Lui2014teichmuller} and applied for the analysis of the cerebral cortex~\cite{Choi2015flash}, hippocampus~\cite{Chan2016quasi}, colon~\cite{Zeng2010supine}, vertebral bone~\cite{Lam2015landmark}, vestibular system~\cite{Wen2015landmark}, human face~\cite{Meng2016tempo}, insect wing~\cite{Jones2013planar,Choi2018planar}, teeth~\cite{Choi2020tooth} etc. However, in the above-mentioned works, the shapes are always assumed to be with a global 1-1 correspondence. In other words, every point on one shape is assumed to correspond uniquely to a point on another shape. For instance, in~\cite{Choi2020tooth}, the occlusal surfaces of the teeth were manually delineated by dental experts, and the quasi-conformal mappings between the teeth assumed a 1-1 correspondence between the segmented tooth boundaries. This assumption limits the use of the methods for general datasets, for which it may be difficult or even impossible to assume a global 1-1 correspondence between every pair of subjects, as there may be errors in the steps of data acquisition, surface reconstruction or manual delineation, such that some parts of one preprocessed surface in fact do not correspond to any position of another preprocessed surface. Therefore, in general, it is more desirable to have a landmark-matching mapping method which is capable of finding an optimal registration between two surfaces without any assumption of a complete surface correspondence. Shape analysis can then be carried out based on the optimal registration results without being affected by the inconsistent parts.

In this work, we propose a new framework for shape analysis based on a recent algorithm for inconsistent surface registration~\cite{Qiu2019inconsistent}. Unlike the prior quasi-conformal mapping methods, the inconsistent surface registration method does not assume any global correspondence between two shapes, making it particularly suitable for tackling general shape analysis problems. Given two inconsistent surfaces with prescribed landmark constraints, we first apply the method to find an optimal landmark-matching quasi-conformal mapping between them. This gives a 1-1 correspondence between the common regions of the two surfaces, which are automatically determined by the method. The remaining inconsistent regions of the two surfaces are not aligned, and hence the quantification of the shape variation of two surfaces can be focused on the common regions without being affected by the inconsistency of the overall shapes. In particular, we evaluate the quasi-conformal distortion, mean curvature difference and Gaussian curvature difference between the common regions of two shapes and use them to construct a dissimilarity measure between the two shapes. Based on the dissimilarity measure between every pair of shapes, we can then perform a cluster analysis to study the shape variation. We demonstrate the effectiveness of our proposed framework by using it to analyze a set of mammalian teeth~\cite{Winchester2014dental,Gao2015hypoelliptic}.

The rest of the paper is organized as follows. In Section~\ref{sect:background}, we review some mathematical concepts relevant to our method. In Section~\ref{sect:main}, we describe our proposed framework for shape analysis via inconsistent surface registration. In Section~\ref{sect:results}, we describe the tooth dataset used in our work and present our results. We conclude our work and discuss future directions in Section~\ref{sect:conclusion}.

\section{Mathematical background}\label{sect:background}

\begin{figure}[t!]
\centering
\includegraphics[width=0.7\textwidth]{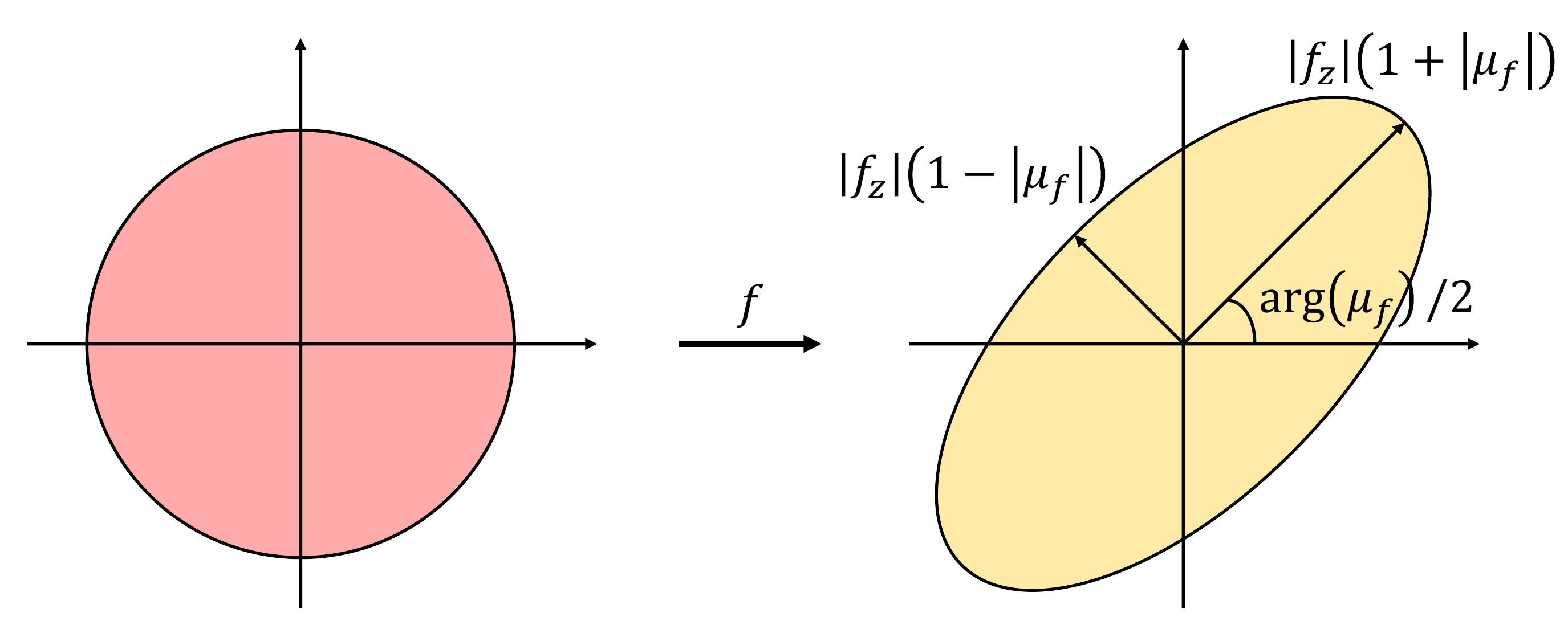}
\caption{Infinitesimal circles are mapped to infinitesimal ellipses with bounded eccentricity under any quasi-conformal map $f$. The maximum magnification, the maximum shrinkage, and the orientation change of the infinitesimal ellipses are $|f_z|(1+|\mu_f|)$, $|f_z|(1-|\mu_f|)$, and $\text{arg}(\mu_f)/2$ respectively.}
\label{fig:qc_figure}
\end{figure}

\subsection{Quasi-conformal theory}
Here we review the basics of quasi-conformal theory and refer the readers to~\cite{Gardiner2000quasiconformal} for more details. Conformal maps preserve angles and hence the local geometry, and quasi-conformal maps are a generalization of conformal maps. Intuitively, conformal maps send infinitesimal circles to infinitesimal circles, while quasi-conformal maps send infinitesimal circles to infinitesimal ellipses with bounded eccentricity. Mathematically, a map $f:\mathbb{C} \to \mathbb{C}$ is said to be a \emph{quasi-conformal map} if it satisfies the Beltrami equation
\begin{equation}\label{eqt:beltrami_equation}
\frac{\partial f}{\partial \bar{z}} = \mu_f(z) \frac{\partial f}{\partial z},
\end{equation}
for some complex-valued function $\mu_f(z)$ with $\|\mu_f(z)\|_{\infty} < 1$. $\mu_f$ is called the \emph{Beltrami coefficient} of $f$, which captures the quasi-conformal distortion of $f$ (see Figure~\ref{fig:qc_figure}). In particular, the maximum magnification and the maximum shrinkage of the infinitesimal ellipses are given by $|f_z|(1+|\mu_f|)$ and $|f_z|(1-|\mu_f|)$ respectively, and hence the aspect ratio of the ellipses, also known as the dilatation, is given by $\frac{1+|\mu_f|}{1-|\mu_f|}$. It is noteworthy that if $\mu_f \equiv 0$ then the right hand side of Eq.~\eqref{eqt:beltrami_equation} becomes 0, which implies that $f$ is conformal. 

Quasi-conformal maps can also be defined between two Riemann surfaces $S_1, S_2$ in $\mathbb{R}^3$ with the aid of local charts. In our work, $S_1$ and $S_2$ are considered to the simply-connected open surfaces. Therefore, if a mapping $f:S_1 \to S_2$ can be decomposed into $f = \psi^{-1} \circ h \circ \phi$, where $\phi: S_1 \to \mathbb{C}$ and $\psi: S_2 \to \mathbb{C}$ are two conformal maps and $h$ is a quasi-conformal map on the complex plane, then the quasi-conformal distortion of $f$ can be represented by the quasi-conformal distortion of $h$. 

\subsection{Differential geometry of surfaces}
In our work, curvatures are used for quantifying shape difference. Here we briefly review the theory of surface curvatures and refer the readers to~\cite{Docarmo1976differential} for details. Let $S$ be a smooth surface in $\mathbb{R}^3$, and let $p$ be a point in $S$. Denote the surface normal at $p$ by $N(p)$. Any normal plane containing $N(p)$ cuts the surface $S$ in a plane curve. One can then evaluate the curvature of the plane curve at $p$. Considering all possible normal planes at $p$, the principal curvatures $\kappa_1(p)$ and $\kappa_2(p)$ are the maximum and minimum values of the curvature of the resulting plane curve. 

Using the principal curvatures, one can obtain the mean curvature $H$ at $p$, which is given by the average of the principal curvatures:
\begin{equation}
H(p) = \frac{1}{2} (\kappa_1(p)+\kappa_2(p)).
\end{equation}
Note that the mean curvature is an extrinsic measure of curvature which depends on the embedding of the surface in the ambient space. One can also obtain the Gaussian curvature $K$ at $p$, which is given by the product of the principal curvatures:
\begin{equation}
K(p) = \kappa_1(p) \kappa_2(p).
\end{equation}
Unlike the mean curvature, the Gaussian curvature is an intrinsic measure of curvature which is independent of the embedding of the surface in the ambient space. 

\begin{figure}[t!]
\centering
\includegraphics[width=\textwidth]{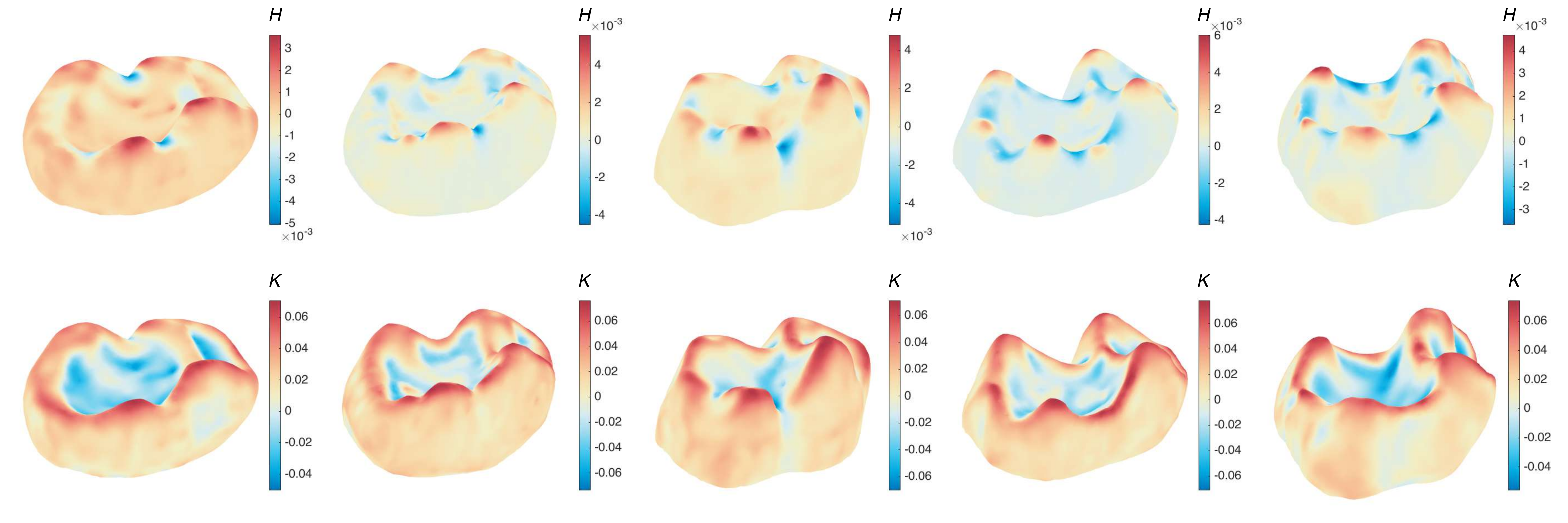}
\caption{Five example tooth surfaces in the mammalian molar dataset~\cite{Winchester2014dental,Gao2015hypoelliptic}, color-coded by the mean curvature $H$ and the Gaussian curvature $K$. Each column corresponds to one tooth surface.}
\label{fig:teeth_curvature}
\end{figure}

Figure~\ref{fig:teeth_curvature} shows the mean and Gaussian curvatures of five tooth surfaces. Altogether, the mean and Gaussian curvatures capture both the extrinsic and intrinsic geometry of any smooth surface in $\mathbb{R}^3$ and hence are useful for shape analysis.

\section{Proposed shape analysis framework} \label{sect:main}
In this section, we describe our proposed shape analysis framework via inconsistent surface registration. The framework is outlined in Algorithm~\ref{alg:framework}, and the details of each step are given in the following sections.

\begin{algorithm}[h]
\KwIn{A set of simply-connected open surfaces $\{S_i\}_{i=1}^n$ with corresponding landmarks.}
\KwOut{$n \times n$ pairwise landmark-matching mapping results between the optimal subdomains, an $n \times n$ dissimilarity matrix $D$, and clustering labels ${l_i}_{i=1}^n$.}
\BlankLine

  \For{$i\gets1$ \KwTo $n$}{
    \For{$j\gets1$ \KwTo $n$}{
        Compute the landmark-matching optimal registration $f_{ij}:\Omega_i \subset S_i \to \Omega_j \subset S_j$, where $\Omega_i$, $\Omega_j$ are the optimal subdomains of $S_i$ and $S_j$ determined by the inconsistent surface registration method (Section~\ref{sect:registration})\;
        }
    }
    
    Using the registration results, obtain an $n\times n$ dissimilarity matrix $D$ with $D(i,j)$ capturing the quasi-conformal distortion, normalized mean curvature difference and normalized Gaussian curvature difference between $S_i$ and $S_j$ (Section~\ref{sect:dissimilarity})\;
    
    Using the dissimilarity matrix $D$, perform a clustering analysis and obtain the clustering labels ${l_i}_{i=1}^n$ (Section~\ref{sect:clustering})\;
    
\caption{Shape analysis via inconsistent surface registration}
\label{alg:framework}
\end{algorithm}

\subsection{Inconsistent surface registration} \label{sect:registration}
\subsubsection{Overview}
\begin{figure}[t!]
\centering
\includegraphics[width=\textwidth]{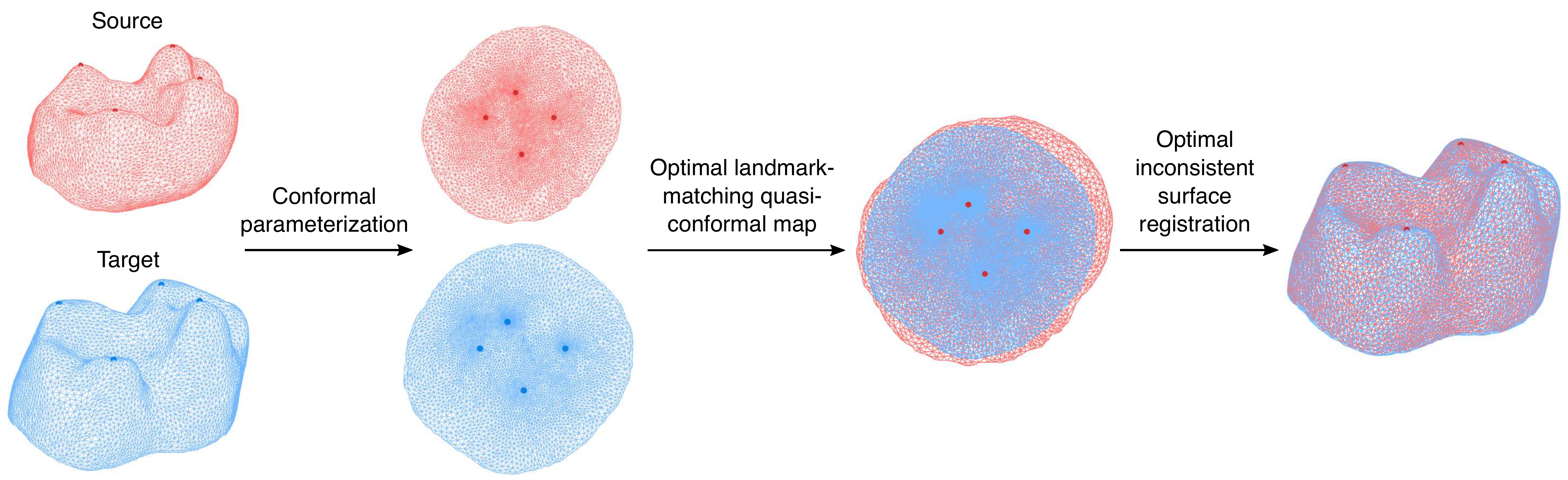}
\caption{The procedure of the inconsistent surface registration. Given two surfaces $S_1, S_2$ with landmark constraints, we first conformally parameterize the two surfaces onto the plane. Then, we find an optimal landmark-matching quasi-conformal map $h:X \to X'$ between two optimal subdomains (the common regions) on the two flattened shapes. Finally, we obtain the desired inconsistent registration $f:\Omega_1 \to \Omega_2$ where $\Omega_1 \subset S_1$ and $\Omega_2 \subset S_2$ are the two optimal subdomains that correspond to $X$ and $X'$ respectively.}
\label{fig:framework}
\end{figure}

Let $S_1, S_2$ be two simply-connected open surfaces in $\mathbb{R}^3$, with corresponding landmarks $\{p_i\}_{i=1}^k$ and $\{q_i\}_{i=1}^k$. We aim to find two optimal common regions $\Omega_1 \subset S_1$ and $\Omega_2 \subset S_2$ and an optimal mapping $f:\Omega_1 \to \Omega_2$, such that $f$ is a bijective landmark-matching mapping satisfying
\begin{equation}
f(p_i) = q_i, i = 1, 2, \dots, k.
\end{equation}
In other words, $\Omega_1$ and $\Omega_2$ are the common regions on the two surfaces with a 1-1 correspondence established by $f$. They contain the landmarks $\{p_i\}_{i=1}^k$ and $\{q_i\}_{i=1}^k$ and capture the key shape variation between the two surfaces. The remaining regions $S_1 \setminus \Omega_1$ and $S_2 \setminus \Omega_2$ are the inconsistent parts of the two surfaces, which will not be taken into account for quantifying the shape variation of the two surfaces. Furthermore, $f$ minimizes the geometric distortion between $\Omega_1$ and $\Omega_2$ in terms of the norm of the Beltrami coefficient $|\mu|$, which provides us with a measure of quasi-conformal dissimilarity of the two common regions $\Omega_1$ and $\Omega_2$. The procedure of the registration is illustrated in Figure~\ref{fig:framework}.

\subsubsection{Conformal parameterization}
Following the approach in~\cite{Qiu2019inconsistent}, we start with flattening the two surfaces $S_1$ and $S_2$ using conformal parameterization. This effectively simplifies the registration problem as we then only need to handle a planar registration problem. In particular, we compute two free-boundary conformal parameterizations $\phi:S_1 \to \mathbb{C}$ and $\psi:S_2 \to \mathbb{C}$. 

Two major advantages of using free-boundary conformal parameterizations are as follows. First, when compared to parameterizations onto a prescribed shape such as the unit disk or a rectangle, free-boundary parameterizations in general have a lower conformal distortion due to the larger flexibility in the flattened shape. Consequently, the discrepancy between the planar registration result and the surface registration result can be reduced. Second, flattening the surfaces onto a standardized shape assumes that the surfaces have a global 1-1 correspondence, which is established via the standardized planar shape. However, in our problem, we do not assume any global 1-1 correspondence between the input surfaces. It is therefore more natural to use free-boundary parameterizations.

Among the existing free-boundary conformal parameterization methods, the least-square conformal map (LSCM) method~\cite{Levy2002least} is used in our framework. LSCM produces a free-boundary conformal flattening map by solving the following energy minimization problem: 
\begin{equation}
    \min_{u,v} \int \frac{1}{2} \|\nabla u\|^2 + \frac{1}{2} \|\nabla v\|^2 - \nabla u \cdot \nabla v^{\perp},
\end{equation}
where $u$ and $v$ are the coordinate functions. More details of the problem formulation and the computational procedure can be found in~\cite{Levy2002least}.

\subsubsection{Optimal planar registration}
After obtaining the conformally flattened domains $\phi(S_1)$ and $\psi(S_2)$, the next step is to find an optimal landmark-matching quasi-conformal map $h:\phi(S_1) \to \mathbb{C}$ with
\begin{equation}
h(\phi(p_i)) = \psi(q_i), i = 1, 2, \dots, k.
\end{equation}
There are infinitely many quasi-conformal mappings that satisfy the landmark-matching constraints. Among them, we search for two optimal subdomains $X\subset \phi(S_1)$ and $X' \subset \psi(S_2)$ and a 1-1 mapping $h:X \to X'$ that minimize the following energy (see~\cite{Qiu2019inconsistent} for more details):
\begin{equation}
E(X, X',h) = \int_{h(X) \cup X'} (I_1^h - I_2)^2 + \int_{X} \left(\lambda|\mu_h|^2 + |\nabla \mu_h|^2\right),
\end{equation}
subject to the landmark constraints $h(\phi(p_i)) = \psi(q_i), i = 1, 2, \dots, k$. Here, $I_1: \phi(S_1) \to \mathbb{R}$ and $I_2:\psi(S_2) \to \mathbb{R}$ are some matching intensities to be prescribed, and $I_1^h := I_1 \circ h^{-1}$ is the deformed image of $I_1$ under $h$. $\mu_h$ is the Beltrami coefficient of $h$. The first integral aims at establishing a meaningful pointwise correspondence by minimizing the shape mismatch error between the two subdomains $X$ and $X'$ in terms of the prescribed intensities, and the second integral is a regularization term that minimizes the local geometric distortion (measured by $|\mu_h|$) and promotes the smoothness of the mapping (measured by $|\nabla \mu_h|$). In practice, we set the matching intensities $I_1, I_2$ based on the Gaussian curvature of the two surfaces, with an appropriate normalization such that $0 \leq I_1, I_2 \leq 1$: 
\begin{equation}
I_1 = \frac{K_1 - \min K_1}{\max K_1 - \min K_1}, \ \ I_2 = \frac{K_2 - \min K_2}{\max K_2 - \min K_2}.
\end{equation} 

A splitting scheme was used for solving the above optimization problem iteratively in~\cite{Qiu2019inconsistent}, which involves alternately solving the landmark-matching quasi-conformal mapping problem and the intensity-based registration problem. In this work, we modify the splitting scheme by first solving the intensity-based registration problem using the Demons method~\cite{Thirion1998image,Pennec1999understanding}, and then the landmark-matching quasi-conformal mapping problem using the Linear Beltrami Solver~\cite{Lui2013texture} at each iteration. This ensures that the intensity mismatch error and quasi-conformal distortion decrease throughout the iterations. Moreover, as our modified splitting scheme always ends with a landmark-matching step, the prescribed landmarks are exactly matched, thereby producing an optimal landmark-matching inconsistent planar registration. In practice, we set the maximum number of iterations to be $n_{\text{iter}} = 20$. The algorithm is outlined in Algorithm~\ref{alg:registration}. 

\begin{algorithm}[h]
\KwIn{Two conformally flattened domains $\phi(S_1)$ and $\psi(S_2)$ of two simply-connected open surfaces $S_1, S_2$ with corresponding landmarks $\{p_i\}_{i=1}^k$ and $\{q_i\}_{i=1}^k$.}
\KwOut{An optimal landmark-matching quasi-conformal map $h:X \to X'$, with $X\subset \phi(S_1)$ and $X' \subset \psi(S_2)$.}
\BlankLine
Set $t = 0$\;
Initialize $h$ to be the identity map\;
\If{$\sum_i \|h(\phi(p_i)) - \psi(q_i)\|_2 > 0$ or $\int |I_1^h - I_2| > 0$}{
  \For{$t\gets1$ \KwTo $n_{\text{iter}}$}{
       Update $h$ by solving the intensity-matching registration problem using the Demons method~\cite{Thirion1998image,Pennec1999understanding};
       
       Update $h$ by solve the landmark-matching quasi-conformal mapping problem using the Linear Beltrami Solver~\cite{Lui2013texture};
       
    }
}
    
\caption{Inconsistent planar registration}
\label{alg:registration}
\end{algorithm}

\subsubsection{Final inconsistent shape registration between surfaces}
After finding the two optimal subdomains $X$ and $X'$ and the optimal landmark-matching quasi-conformal map $h:X\to X'$, we can obtain two corresponding common regions $\Omega_1 \subset S_1$ and $\Omega_2 \subset S_2$ by $\Omega_1 = \phi^{-1}(X)$ and $\Omega_2 = \psi^{-1}(X')$. It is easy to see that all corresponding landmarks in $S_1$ and $S_2$ are guaranteed to be in $\Omega_1$ and $\Omega_2$, and they are exactly matched under $h$.

Finally, the landmark-matching inconsistent shape registration mapping $f:\Omega_1 \to \Omega_2$ is given by
\begin{equation}
f = \psi^{-1} \circ h \circ \phi.
\end{equation}
Since $\phi$ and $\psi$ are conformal, the quasi-conformal distortion of $f$ is equal to that of $h$.

\subsection{Quantifying the dissimilarity between inconsistent shapes} \label{sect:dissimilarity}
For any simply-connected open surfaces $S_1, S_2$, using the method introduced in Section~\ref{sect:registration}, we obtain a landmark-matching quasi-conformal map $f:\Omega_1 \subset S_1 \to \Omega_2 \subset S_2$. We can then quantify the difference between $S_1$ and $S_2$ using the correspondence between $\Omega_1$ and $\Omega_2$. More specifically, three important shape indices can be considered: 
\begin{enumerate}[(i)]
    \item The quasi-conformal distortion $\mu_f$ of the mapping $f$, which captures the effort needed to deform $\Omega_1$ to match $\Omega_2$ in terms of the eccentricity of the infinitesimal circle packing. Note that by quasi-conformal theory, for any quasi-conformal map $f$, we always have
    \begin{equation}
        0 \leq |\mu_f| \leq 1.
    \end{equation}
    \item The normalized mean curvature difference $\frac{1}{2} |\tilde{H}_1(f) - \tilde{H}_2|$ under the optimal registration $f$, where $\tilde{H}_1, \tilde{H}_2$ are the normalized mean curvature difference of $S_1$ and $S_2$ with $0 \leq |\tilde{H}_1|, |\tilde{H}_2| \leq 1$. Note that with the normalization, we always have $|\tilde{H}_1(f) - \tilde{H}_2| \leq 1 - (-1) = 2$ and hence 
    \begin{equation}
        0 \leq \frac{1}{2} |\tilde{H}_1(f) - \tilde{H}_2| \leq 1.
    \end{equation}
    \item The normalized Gaussian curvature difference $\frac{1}{2} |\tilde{K}_1(f) - \tilde{K}_2|$ optimal registration $f$, where $\tilde{K}_1, \tilde{K}_2$ are the normalized Gaussian curvature difference of $S_1$ and $S_2$ with $0 \leq |\tilde{K}_1|, |\tilde{K}_2| \leq 1$. Again, with the normalization, we always have $|\tilde{K}_1(f) - \tilde{K}_2| \leq 1 - (-1) = 2$ and hence 
    \begin{equation}
        0 \leq \frac{1}{2} |\tilde{K}_1(f) - \tilde{K}_2| \leq 1.
    \end{equation}
\end{enumerate}

Now, we define the combined shape index $\delta(S_1,S_2)$ as follows:
\begin{equation}
    \delta(S_1,S_2) = \frac{1}{A(\Omega_1)} \int_{\Omega_1} \left( \alpha |\mu_f| + \frac{\beta}{2} |\tilde{H}_1(f) - \tilde{H}_2| + \frac{\gamma}{2} |\tilde{K}_1(f) - \tilde{K}_2| \right),
\end{equation}
where $A(\Omega_1)$ is the area of the optimal region $\Omega_1$, and $\alpha, \beta, \gamma \geq 0$ are the weighting factors for the three quantities with $\alpha+\beta+\gamma = 1$. It follows that
\begin{equation}
    0 \leq \delta(S_1,S_2) \leq \frac{1}{A(\Omega_1)} \int_{\Omega_1} \left( \alpha + \beta + \gamma \right) \leq \frac{1}{A(\Omega_1)} \int_{\Omega_1} 1 = 1.
\end{equation}
Alternatively, one can obtain a landmark-matching quasi-conformal map $g:\Omega_2 \subset S_2 \to \Omega_1 \subset S_1$. This gives us
\begin{equation}
    \delta(S_2,S_1) = \frac{1}{A(\Omega_2)} \int_{\Omega_2} \left( \alpha |\mu_g| + \frac{\beta}{2} |\tilde{H}_2(g) - \tilde{H}_1|+ \frac{\gamma}{2} |\tilde{K}_2(g) - \tilde{K}_1|  \right),
\end{equation}
and similarly we have $0 \leq \delta(S_2,S_1) \leq 1$. Based on the above quantities, we define the dissimilarity $d$ between two inconsistent surfaces $S_1$ and $S_2$ by
\begin{equation}
    d(S_1, S_2) = \min \{\delta(S_1,S_2), \delta(S_2,S_1)\} \in [0,1].
\end{equation}
Note that a dissimilarity matrix for $n$ shapes is an $n \times n$ matrix satisfying the following two properties:
\begin{enumerate}[(i)]
    \item The matrix is symmetric.
    \item All diagonal entries of the matrix should be zero, as there is no dissimilarity between a shape and itself.
\end{enumerate}
Now, after computing the pairwise mappings between all $n$ surfaces $\{S_i\}_{i=1}^n$, we define an $n \times n$ matrix $D$ by $D(i,j) = d(S_i, S_j)$. By the definition of $d$, we have 
\begin{equation}
    D(i,j) = d(S_i,S_j) = \min \{\delta(S_i,S_j), \delta(S_j,S_i)\} = d(S_j, S_i) = M(j,i)
\end{equation}
for all $1\leq i,j \leq n$. Also, since the identity map $I:S_i \to S_i$ is conformal, we have
\begin{equation}
    D(i,i) = d(S_i,S_i) = \frac{1}{A(\Omega)} \int_{\Omega} \left( \alpha \cdot 0 + \frac{\beta}{2} |\tilde{H}_i - \tilde{H}_i| + \frac{\gamma}{2} |\tilde{K}_i - \tilde{K}_i| \right) = 0,
\end{equation}
for any $\alpha, \beta, \gamma$. Therefore, $D$ satisfies both conditions (i) and (ii) and is a valid dissimilarity matrix.

\subsection{Cluster analysis based on geometric difference} \label{sect:clustering}
After obtaining the dissimilarity matrix $D$ that captures the shape difference of the $n$ surfaces $\{S_i\}_{i=1}^n$ in terms of the quasi-conformal distortion, the normalized mean curvature difference and the normalized Gaussian curvature difference, we can then cluster the surfaces into different groups according to their pairwise geometric dissimilarity with the aid of clustering algorithms. 

Cluster analysis is a well-studied topic in statistics, and many useful algorithms have been proposed. In this work, we consider two widely-used clustering methods, namely the hierarchical clustering method and the k-means clustering method. Below, we briefly describe the two methods and refer the readers to~\cite{Rokach2005clustering} for more details.

The hierarchical clustering method starts by linking pairs of surfaces which are close together into binary clusters, i.e. clusters with exactly two objects, based on the dissimilarity measure encoded in $D$. It then continues to link these binary clusters with each other to form bigger clusters. Ultimately, all $n$ objects are linked together in a binary hierarchical cluster tree. Then, one can cluster all $n$ surfaces into a certain number of clusters by cutting off the hierarchy at different levels in the binary tree. 

The k-means clustering method aims to cluster $n$ observations into $k$ clusters, such that each observation belongs to the cluster with the nearest mean. Each observation is required to be a $p$-dimensional coordinate vector. Starting from our dissimilarity matrix $D$, we can first apply the multidimensional scaling (MDS) method to construct an optimal projection of the $n$ shapes onto a $p$-dimensional Euclidean space, so that each shape $S_i$ is represented by a data point ${\bf x}_i \in \mathbb{R}^p$. In particular, MDS aims to minimize the difference between $\|{\bf x}_i - {\bf x}_j\|$ and $D(i,j)$, so that the information of the pairwise distances given by $D$ is optimally translated into a set of coordinates. Now, given the $n$ points $\{{\bf x}_i\}_{i=1}^n$ and a prescribed positive integer $k$, the k-means clustering method looks for an optimal grouping of the points into $k$ clusters $C_1, C_2, \dots, C_k$, such that the sum of the squared distance between each data point and the center of its corresponding cluster, i.e. $\sum_{j=1}^k \sum_{{\bf x} \in C_j} \| {\bf x} - {\bm \mu}_j\|^2$, is minimized. Here, ${\bm \mu}_j$ is the mean of the points in $C_j$.

\section{Experiment} \label{sect:results}

\begin{table}[t!]
\footnotesize
 \centering
 \begin{tabular}{|c|c|c|} \hline
  Genus & Species & Specimen ID \\ \hline
  
  \multirow{10}{*}{\emph{Alouatta}} & \emph{Alouatta seniculus} & 32 \\ \cline{2-3}
  & \emph{Alouatta seniculus} & 33 \\ \cline{2-3}
  & \emph{Alouatta palliata} & 34 \\ \cline{2-3}
  & \emph{Alouatta seniculus} & 36 \\ \cline{2-3}
  & \emph{Alouatta seniculus} & 37 \\ \cline{2-3}
  & \emph{Alouatta seniculus} & 38 \\ \cline{2-3}
  & \emph{Alouatta seniculus} & 39 \\ \cline{2-3}
  & \emph{Alouatta seniculus} & 40 \\ \cline{2-3}
  & \emph{Alouatta palliata} & 41 \\ \cline{2-3}
  & \emph{Alouatta palliata} & 42 \\ \hline
  
  \multirow{10}{*}{\emph{Ateles}} & \emph{Ateles belzebuth} & 3 \\ \cline{2-3}
  & \emph{Ateles belzebuth} & 4 \\ \cline{2-3}
  & \emph{Ateles belzebuth} & 9 \\ \cline{2-3}
  & \emph{Ateles geoffroyi} & 20 \\ \cline{2-3}
  & \emph{Ateles paniscus} & 21 \\ \cline{2-3}
  & \emph{Ateles geoffroyi} & 22 \\ \cline{2-3}
  & \emph{Ateles belzebuth} & 35 \\ \cline{2-3}
  & \emph{Ateles geoffroyi} & 43 \\ \cline{2-3}
  & \emph{Ateles belzebuth} & 46 \\ \cline{2-3}
  & \emph{Ateles belzebuth} & 47 \\ \hline
  
  \multirow{10}{*}{\emph{Brachyteles}} & \emph{Brachyteles arachnoides} & 19\\ \cline{2-3}
  & \emph{Brachyteles arachnoides} & 23 \\ \cline{2-3}
  & \emph{Brachyteles arachnoides} & 24 \\ \cline{2-3}
  & \emph{Brachyteles arachnoides} & 25 \\ \cline{2-3}
  & \emph{Brachyteles arachnoides} & 26 \\ \cline{2-3}
  & \emph{Brachyteles arachnoides} & 27 \\ \cline{2-3}
  & \emph{Brachyteles arachnoides} & 28 \\ \cline{2-3}
  & \emph{Brachyteles arachnoides} & 29 \\ \cline{2-3}
  & \emph{Brachyteles arachnoides} & 30 \\ \cline{2-3}
  & \emph{Brachyteles arachnoides} & 31 \\ \hline
  
  \multirow{10}{*}{\emph{Callicebus}} & \emph{Callicebus donacophilus} & 2 \\ \cline{2-3}
  & \emph{Callicebus torquatus} & 8\\ \cline{2-3}
  & \emph{Callicebus torquatus} & 10\\ \cline{2-3}
  & \emph{Callicebus torquatus} & 11\\ \cline{2-3}
  & \emph{Callicebus torquatus} & 12\\ \cline{2-3}
  & \emph{Callicebus torquatus} & 13\\ \cline{2-3}
  & \emph{Callicebus moloch} & 14\\ \cline{2-3}
  & \emph{Callicebus moloch} & 15\\ \cline{2-3}
  & \emph{Callicebus donacophilus} & 17\\ \cline{2-3}
  & \emph{Callicebus donacophilus} & 18\\ \hline
  
  \multirow{10}{*}{\emph{Saimiri}} & \emph{Saimiri boliviensis} & 1 \\ \cline{2-3}
  & \emph{Saimiri boliviensis} & 5 \\ \cline{2-3}
  & \emph{Saimiri boliviensis} & 6 \\ \cline{2-3}
  & \emph{Saimiri boliviensis} & 7 \\ \cline{2-3}
  & \emph{Saimiri boliviensis} & 16\\ \cline{2-3}
  & \emph{Saimiri boliviensis} & 44 \\ \cline{2-3}
  & \emph{Saimiri boliviensis} & 45 \\ \cline{2-3}
  & \emph{Saimiri sciureus} & 48 \\ \cline{2-3}
  & \emph{Saimiri sciureus} & 49 \\ \cline{2-3}
  & \emph{Saimiri sciureus} & 50 \\ \hline
 \end{tabular}
\caption{The list of the 50 tooth specimens adapted from the Platyrrhine molar dataset~\cite{Winchester2014dental,Gao2015hypoelliptic}.} 
\label{table:list}
\end{table}

\begin{figure}[t!]
\centering
\includegraphics[width=0.75\textwidth]{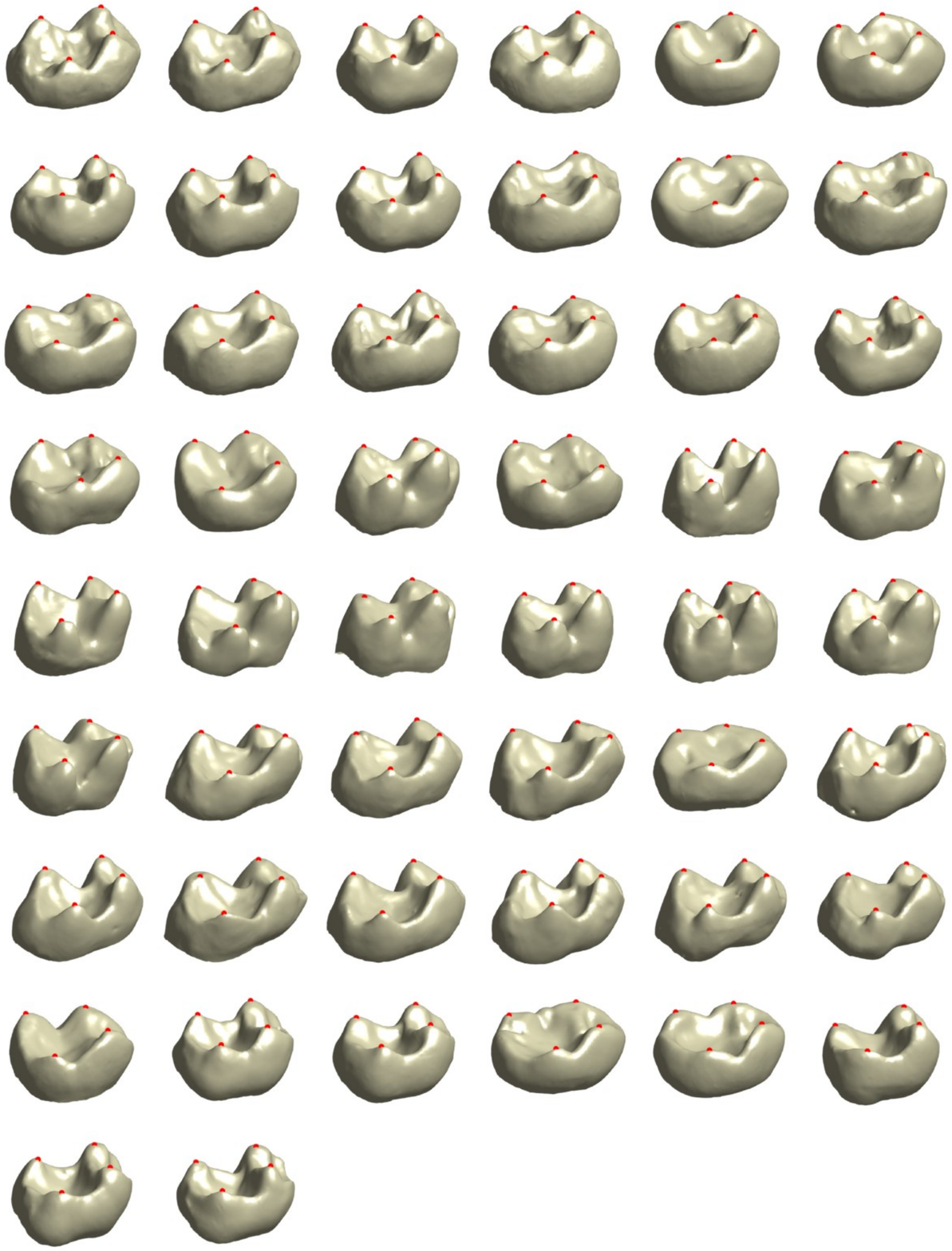}
\caption{The 50 tooth specimens in the Platyrrhine molar dataset~\cite{Winchester2014dental,Gao2015hypoelliptic}, each with four prescribed landmarks at the four cusps. The specimens are plotted according to the specimen IDs in Table~\ref{table:list}, with the first row showing specimens 1--6, and the second row showing specimens 7--12 etc.}
\label{fig:teeth_all}
\end{figure}
\subsection{Dataset}
To demonstrate the effectiveness of our proposed shape analysis framework, we study the Platyrrhine (New World monkey) molar dataset~\cite{Winchester2014dental,Gao2015hypoelliptic}, which consists of 50 second mandibular molar specimens of primates publicly available at the biological data archive \emph{MorphoSource}~\cite{Morphosource}. The specimens were obtained from the American Museum of Natural History, the Smithsonian Institution National Museum of Natural History, the Harvard University Museum of Comparative Zoology, and the Stony Brook University Museum of Anatomy. The 50 molars were evenly collected from 5 genera, namely \emph{Alouatta} (also known as Howler monkeys), \emph{Ateles} (also known as Spider monkeys), \emph{Brachyteles} (also known as woolly spider monkeys), \emph{Callicebus} (also known as titis), and \emph{Saimiri} (also known as squirrel monkeys). Table~\ref{table:list} records the genus and species of each specimen. The specimens were discretized in the form of triangle meshes, each containing about 5000 vertices and 10000 faces (see~\cite{Winchester2014dental,Gao2015hypoelliptic} for more details of the data acquisition and preprocessing procedure). We further preprocessed the meshes to ensure that they are all simply-connected, open, and without any non-manifold vertices or edges. 

Note that for each molar in the dataset, the most prominent features are the four cusps. To establish a correspondence between the common parts of two molars, such features should be consistently aligned. Therefore, we extracted the four vertices at the apex of the four cusps as landmarks. The landmark-matching inconsistent shape registration method described in the above section ensures that the four landmarks are exactly matched under the mappings, so that a systematic comparison between the molars can be subsequently performed.

Figure~\ref{fig:teeth_curvature} shows five example molar surfaces in the dataset color-coded by the mean and Gaussian curvatures. Figure~\ref{fig:teeth_all} shows all 50 molar surfaces in the dataset, each with the four prescribed landmarks labeled.

\subsection{Implementation}
The proposed framework and the relevant algorithms were implemented in MATLAB. The linear systems were solved using the MATLAB backslash operator ($\backslash$). The computations were performed on a Windows PC with a quad core processor and 16~GB RAM. To accelerate the computations of the $50\times 50 = 2500$ pairwise mappings between all 50 tooth meshes, we used the \texttt{parfor} function in the Parallel Computing Toolbox in MATLAB, with all four CPU cores of the PC utilized. The computation of each registration mapping took around 40 seconds, and the computations of all 2500 pairwise mappings took around eight hours in total.

The cluster analysis was also done using MATLAB. For hierarchical clustering, we used the MATLAB built-in functions \texttt{linkage} and \texttt{cluster}. For multidimensional scaling, we used the MATLAB built-in function \texttt{mdscale}. For k-means clustering, we used the MATLAB built-in function \texttt{kmeans} with the parameter \texttt{Replicates} being 100.

\begin{figure}[t!]
    \centering
    \includegraphics[width=\textwidth]{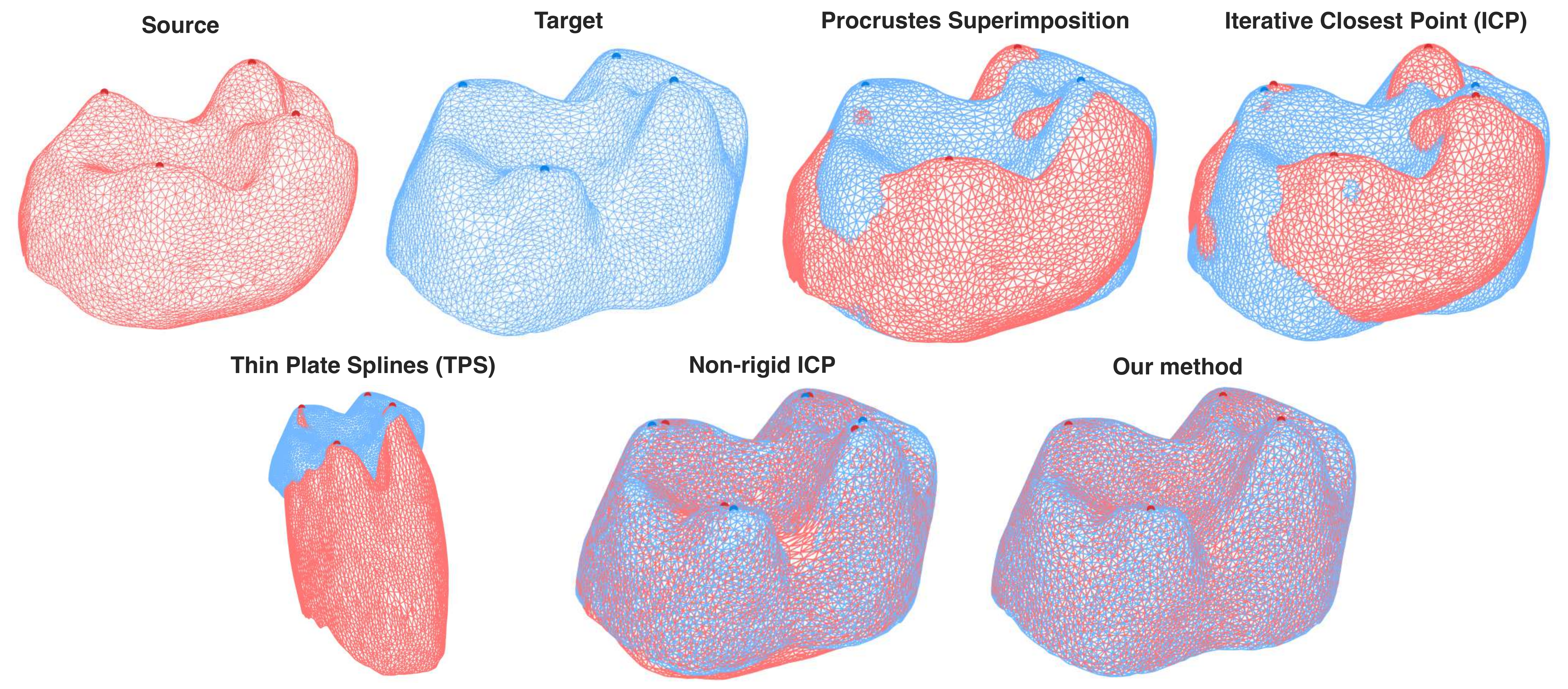}
    \caption{Computing a mapping between two molars. Top row (left to right): the source surface with four landmarks, the target surface with four landmarks, the mappings produced by the Procrustes superimposition method~\cite{Gower1975generalized} and the iterative closest point (ICP) method~\cite{Besl1992a}. Bottom row (left to right): the mappings produced by the thin plate splines (TPS) method~\cite{Bookstein1989principal}, the non-rigid ICP method~\cite{Audenaert2019cascaded}, and our inconsistent surface registration method.}
    \label{fig:pairwise_result}
\end{figure}
\subsection{Pairwise mapping result}
We first consider computing pairwise mappings between the molars using various methods. As shown in Figure~\ref{fig:pairwise_result}, for rigid transformation methods such as the Procrustes superimposition method~\cite{Gower1975generalized} and the iterative closest point (ICP) method~\cite{Besl1992a}, the two sets of corresponding landmarks cannot be exactly matched, and the remaining parts of the two molars are also misaligned. As for non-rigid transformation methods, we observe that while the thin plate splines (TPS) method~\cite{Bookstein1989principal} is able to match the two sets of landmarks, the induced deformation of the source surface is large and hence the overall shapes are largely mismatched. Also, while the non-rigid ICP method~\cite{Audenaert2019cascaded} is capable of deforming the source surface to match the overall geometry of the target surface, the landmarks are not exactly matched. In summary, all of the above methods either suffer from being unable to exactly match the landmarks or to match the surface geometry. In contrast to the above methods, our inconsistent surface registration approach is capable of exactly matching all landmark pairs and appropriately deforming the source surface to match the target surface. This demonstrates the effectiveness of our approach for computing pairwise mappings between surfaces.

\begin{figure}[t!]
    \centering
    \includegraphics[width=\textwidth]{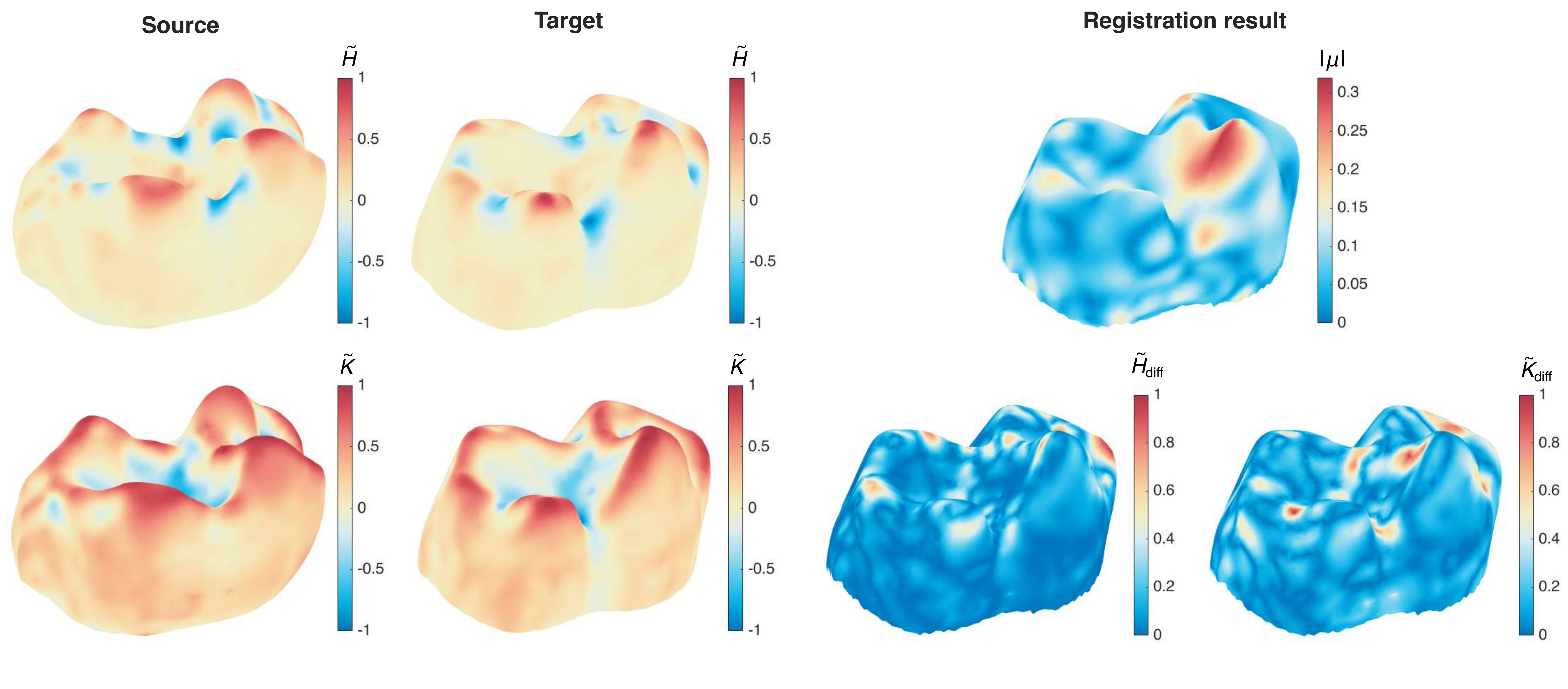}
    \caption{Quantifying the shape difference between two molars using our inconsistent surface registration method. Left column: the normalized mean curvature and the normalized Gaussian curvature of the source molar surface. Middle column: the normalized mean curvature and the normalized Gaussian curvature of the target molar surface. Right column: the norm of the Beltrami coefficients $|\mu|$, the normalized mean curvature difference and the normalized Gaussian curvature difference computed based on the inconsistent surface registration result.}
    \label{fig:pairwise_result_inconsistent}
\end{figure}

From the inconsistent surface registration result, one can easily quantify the difference between two shapes. Specifically, because of the 1-1 correspondence between the optimal subdomains of the source and the target surfaces, we can assess the quasi-conformal distortion (in terms of the norm of the Beltrami coefficients $|\mu|$), the normalized mean curvature difference, and the normalized Gaussian curvature difference between them. A comparison between two molars is shown in Figure~\ref{fig:pairwise_result_inconsistent}. It can be observed that the norm of the Beltrami coefficients $|\mu|$ effectively captures the cusp difference between the two molars. The normalized mean curvature difference, and the normalized Gaussian curvature difference also highlight subtle shape difference between different parts of the two molars. This suggests that our approach is useful for quantifying shape difference.

\begin{figure}[t!]
    \centering
    \includegraphics[width=\textwidth]{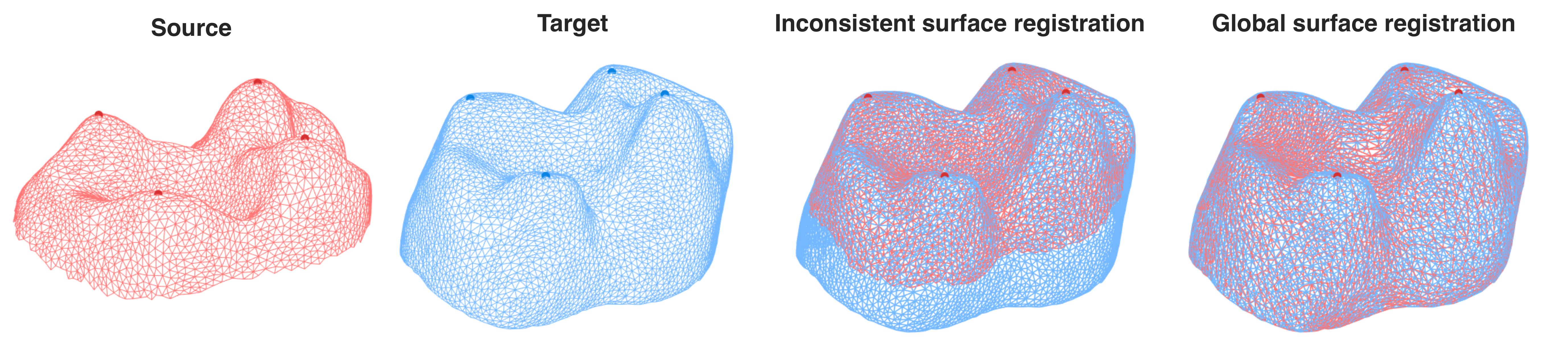}
    \caption{A comparison between our inconsistent surface registration approach and the global landmark-matching quasi-conformal Teichm\"uller mapping method~\cite{Choi2020tooth}. Here, the source mesh (leftmost) and the target mesh (second left) are inconsistently segmented. Our inconsistent surface registration approach successfully identifies and registers the common regions of the two meshes (second right). On the contrary, the landmark-matching Teichm\"uller mapping method~\cite{Choi2020tooth} constructs a global 1-1 correspondence between the two meshes including the inconsistent parts, thereby leading to an overall misalignment with a large distortion (rightmost).}
    \label{fig:pairwise_result_inconsistent_vs_global}
\end{figure}

To highlight the difference between our approach and the previous global landmark-matching quasi-conformal Teichm\"uller mapping method~\cite{Choi2020tooth}, we further consider comparing the two molars that are inconsistently segmented (see Figure~\ref{fig:pairwise_result_inconsistent_vs_global}). For the previous global quasi-conformal mapping method, the inconsistent parts of the two surfaces will be included in the computation. Hence, the global registration is largely distorted and fails to reflect the actual shape correspondence of the two molars. By contrast, our approach prevents such misalignment by optimally detecting the common regions of the two meshes. With the weaker assumption on the global surface correspondence, our approach is applicable to a wider class of shape analysis problems. 

\subsection{Cluster analysis of Platyrrhine molars}
After demonstrating the effectiveness of our method for pairwise mappings, we deploy the method to register all $50\times 50 = 2500$ pairs of the Platyrrhine molars. This provides us with the quasi-conformal distortion, the normalized mean curvature difference, and the normalized Gaussian curvature difference between the optimal subdomains of each pair of molars. We then proceed to use these quantities to cluster the 50 specimens into different groups.

\subsubsection{Binary clustering}
\begin{figure}[t]
    \centering
    \includegraphics[width=\textwidth]{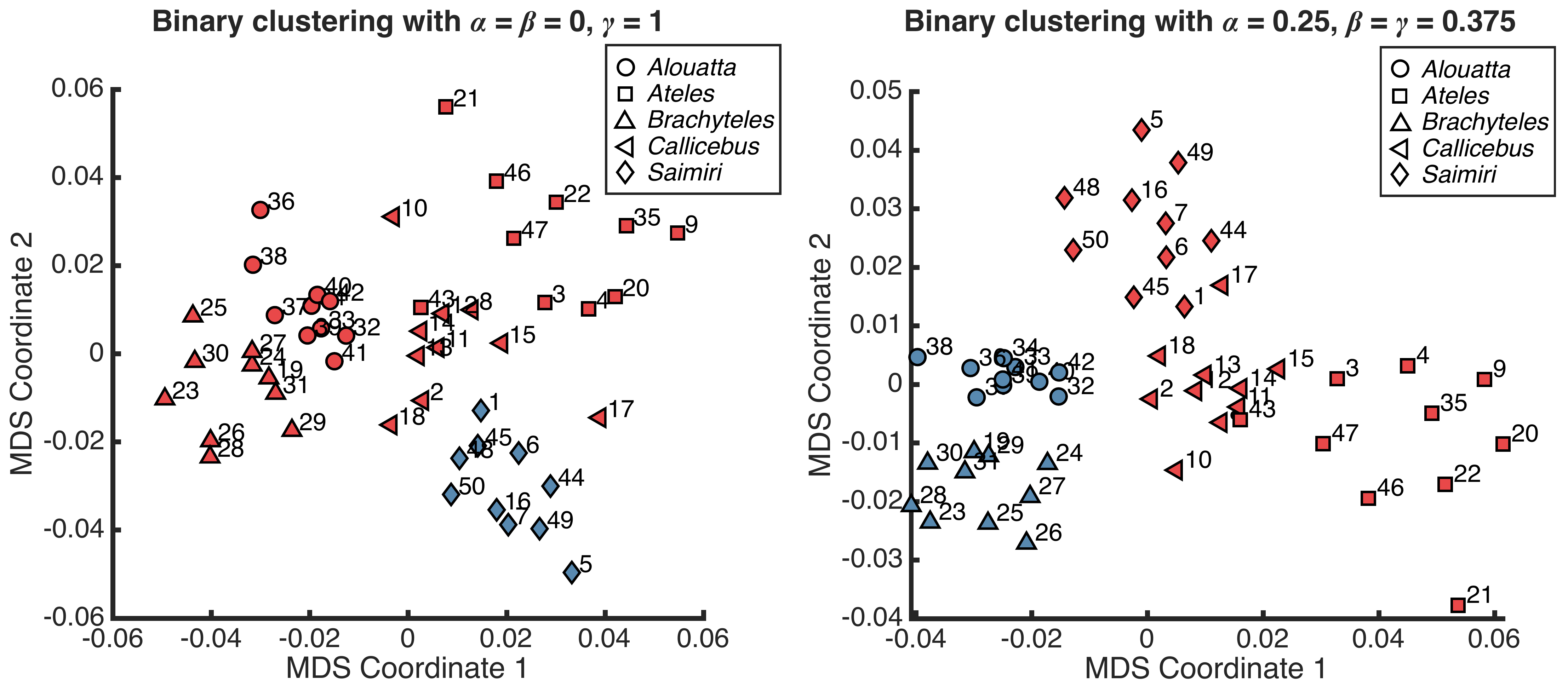}
    \caption{The binary clustering results produced by the hierarchical clustering method using our proposed dissimilarity measure, visualized on the multidimensional scaling (MDS) plane. The specimens from different genera are represented using different markers, and the colors of the markers represent the clustering result. The numerical label of each node corresponds to the specimen ID in Table \ref{table:list}. For the result shown on the left, the parameters $(\alpha, \beta, \gamma) = (0, 0, 1)$ are used for constructing the dissimilarity matrix $D$. One of the resulting clusters consists of all \emph{Saimiri} specimens (insectivores), while the other one consists of all non-insectivores. For the result shown on the right, the parameters $(\alpha, \beta, \gamma) = (0.25, 0.375, 0.375)$ are used for constructing $D$. One of the resulting clusters consists of all \emph{Alouatta} and \emph{Brachyteles} specimens (folivores), while the other one consists of all non-folivores.}
    \label{fig:binary_result}
\end{figure}

We first consider clustering all specimens in the Platyrrhine molar dataset into two groups using the hierarchical clustering method, with some specific choices of the shape index weighting factors $\alpha, \beta, \gamma$ for building the dissimilarity matrix $D$ examined.

If only the quasi-conformal distortion is considered (i.e. $\alpha = 1, \beta = 0, \gamma = 0$) or only the normalized mean curvature difference is considered (i.e. $\alpha = 0, \beta = 1, \gamma = 0$) for constructing the dissimilarity matrix, almost everything is clustered into one group. This suggests that simply using the quasi-conformal distortion or the mean curvature is insufficient to distinguish between the shapes of the Platyrrhine molars. However, for $\alpha = 0, \beta = 0, \gamma = 1$, i.e. only the normalized Gaussian curvature difference is considered, we observe that all 10 \emph{Saimiri} specimens are clustered into one group, while the other 40 specimens are clustered into another group (see Figure~\ref{fig:binary_result} (left)). One may wonder if there is any relationship between the shape of the teeth and their biological functions. Referring to the diets of these species, it is noteworthy that \emph{Saimiri} are insectivores, i.e. animals that eat mainly insects, while the other four genera are not~\cite{primate}. This suggests a potential intrinsic geometric difference between the molars of different genera due to a difference in diet.

More relationships between function and shape can be observed for some other choices of non-zero $\alpha, \beta, \gamma$. For instance, for $(\alpha, \beta, \gamma) = (0.25, 0.375, 0.375)$, all 20 specimens from \emph{Alouatta} and \emph{Brachyteles} are clustered into one group, and the other 30 specimens from \emph{Ateles}, \emph{Callicebus} and \emph{Saimiri} are clustered into another group (see Figure~\ref{fig:binary_result} (right)). Interestingly, \emph{Alouatta} and \emph{Brachyteles} are folivores, i.e. animals that eat primarily leaves, while the species in the other three genera in the dataset are non-folivores~\cite{primate}. According to the theory of primate adaptations~\cite{Fleagle2013primate}, folivores are characterized by molars with extensive development of shearing crests for reducing the leaves into small particles so as to enhance digestion. It can be observed qualitatively from the images in Figure~\ref{fig:teeth_all} that the \emph{Alouatta} and \emph{Brachyteles} teeth have relatively taller cusps and deeper shearing crests when compared with the teeth from the other genera, which is possibly captured by such combination of shape index parameters $\alpha, \beta, \gamma$. Once again, our shape analysis framework sheds light on the interplay between functions and shapes.

\subsubsection{Clustering into five groups}
\begin{figure}[t]
    \centering
    \includegraphics[width=\textwidth]{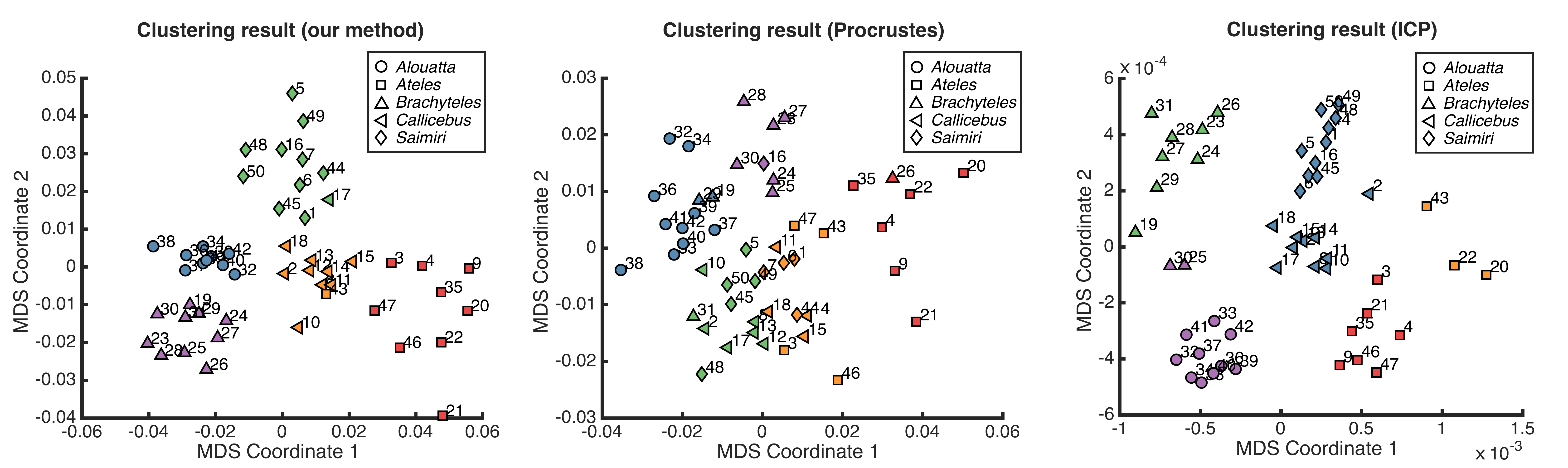}
    \caption{The clustering results produced by the k-means clustering method with $k = 5$ using different dissimilarity measures, visualized on the multidimensional scaling (MDS) plane. The specimens from different genera are represented using different markers, and the colors of the markers represent the clustering result. The numerical label of each node corresponds to the specimen ID in Table \ref{table:list}. Left: the clustering result obtained from our method. Middle: the clustering result obtained from the Procrustes superimposition method~\cite{Gower1975generalized}. Right: the clustering result obtained from the Iterative Closest Point (ICP) method~\cite{Besl1992a}.}
    \label{fig:kmeans_result}
\end{figure}
It is also natural to ask if our method is capable of clustering the molars from different genera into different groups. To answer this question, we consider clustering the entire molar dataset into five groups using the k-means clustering method and evaluating the classification accuracy based on their respective genera. Here, the classification accuracy is defined by the percentage of correct cluster label pairs (i.e. two specimens from the same genus are with the same cluster label, or two specimens from different genera are with two different cluster labels) among all distinct pairs of specimens. The multidimensional scaling (MDS) method is first used to construct a projection of the molars onto the two-dimensional plane. Then, the k-means clustering is performed with $k = 5$.

We vary each of the parameters $\alpha, \beta, \gamma$ in the dissimilarity matrix $D$ from 0 to 1, with an increment of 0.05. The maximum classification accuracy is $96.98\%$, achieved by multiple choices of parameters $(\alpha, \beta, \gamma)$ with $\alpha$ ranging from $0.1$ to $0.25$, $\beta$ ranging from $0.15$ to $0.55$, and $\gamma$ ranging from $0.3$ to $0.6$. Figure~\ref{fig:kmeans_result} (left) shows the classification result for one of the optimal sets of parameters $(\alpha, \beta, \gamma) = (0.2, 0.4, 0.4)$. It can be observed that all specimens from \emph{Alouatta} and \emph{Brachyteles} are correctly clustered, and only one specimen from \emph{Ateles} and one from \emph{Callicebus} are wrongly clustered. 

For comparison, we consider clustering the 50 specimens into five groups using the Procrustes superimposition method~\cite{Gower1975generalized} and the Iterative Closest Point (ICP) method~\cite{Besl1992a}. More specifically, we apply the methods to compute the $50\times 50$ pairwise mappings and use their respective distance measures to form dissimilarity matrices. We then apply the k-means clustering on the dissimilarity matrices and evaluate the classification accuracy. The clustering results are shown in Figure~\ref{fig:kmeans_result} (middle and right). Note that the Procrustes method results in a classification accuracy of $81.47\%$, and the ICP method results in a classification accuracy of $87.18\%$. The better classification achieved by our method can be explained by the fact that the quasi-conformal distortion, the normalized mean curvature difference and the normalized Gaussian curvature difference are all captured by our dissimilarity matrix $D$.

To assess the predictive performance of our clustering method and test if overfitting occurs, we further perform a leave-one-out cross-validation (LOOCV) on our method. Our method achieves a LOOCV accuracy of 94\%, which indicates that it is highly accurate for predicting new data.

Altogether, the experimental results suggest that our proposed framework is capable of revealing the relationship between shape and phylogeny.

\section{Conclusion and future works} \label{sect:conclusion}
In this work, we have proposed a new framework for quantifying shape variation via inconsistent surface registration. Given a set of simply-connected open surfaces with prescribed landmarks, we compute an optimal landmark-matching quasi-conformal map between each pair of surfaces. Instead of assuming a global 1-1 correspondence such that every part of one surface corresponds to a certain part of the other surface, our approach searches for the optimal subdomains on the two surfaces that yield the best correspondence. This prevents the mapping results from being affected by the potential segmentation errors in the data. We then quantify the dissimilarity of the surfaces by evaluating the quasi-conformal distortion, the normalized mean curvature difference, and the normalized Gaussian curvature difference using the correspondence between the optimal subdomains. This further enables us to cluster the shapes based on their geometric difference. We have demonstrated the effectiveness of our proposed framework by deploying it on a mammalian tooth dataset. In particular, unlike the prior shape matching methods, our framework allows for the accurate quantification of shape difference between two tooth surfaces, with the prescribed landmarks exactly matched and the common regions optimally aligned. Our method also successfully produces geometry-based clustering results that highly correlate with the underlying biological difference between the specimens. 

Note that our proposed framework allows for the exact correspondence of feature landmarks in computing the optimal inconsistent shape registration, based on the assumption that the prescribed landmark positions are precise. In case there are potential manual errors in the landmark labeling step so that an inexact landmark-matching registration is more desirable, we can simply add an intensity-matching step at the end of Algorithm~\ref{alg:registration}, thereby producing an optimal landmark-guided registration with a small degree of landmark mismatch allowed. Also, our method can be potentially extended for surfaces with other topology, such as genus-0 closed surfaces and high-genus surfaces. For instance, to handle genus-0 closed surfaces, one may replace the current conformal flattening step with a spherical conformal parameterization step~\cite{Choi2015flash} followed by the stereographic projection. For high-genus surfaces, one may replace the current conformal flattening step with a conformal parameterization onto a fundamental domain on the plane.

In the future, we plan to deploy our inconsistent shape analysis framework to study other biological shapes. In particular, our method is suitable for analyzing human faces as each human face contains prominent landmarks such as the eyes, the nose and the mouth, while it is usually hard to precisely segment the face boundary due to 3D scanning errors and potential hair occlusions etc. We also plan to investigate the combination of our method with other landmark-free anatomical surface mapping methods such as~\cite{Boyer2011algorithms} for further improving the accuracy of shape classification.\\

{\bf Acknowledgments } This work was supported by HKRGC GRF (Reference: 14304715, Project ID: 2130447) (to L.M. Lui). This work was also supported in part by the Harvard Quantitative Biology Initiative and the NSF-Simons Center for Mathematical and Statistical Analysis of Biology at Harvard, award no. 1764269 (to Gary P. T. Choi).

\end{document}